\documentstyle[prb,aps,psfig]{revtex}

\begin{document}

\title{A Microscopic Model for D-Wave Pairing in the Cuprates: What
Happens when Electrons Somersault?  }

\author{Mona Berciu$^{1,2}$ and Sajeev John$^1$}

\address{$^1$Department of Physics, University of Toronto, 60 St. George
Street, Toronto, Ontario, M5S 1A7, Canada}

\address{$^2$Department of Electrical Engineering, Princeton University,
Princeton, New Jersey 08544}

\date{\today}

\maketitle


\begin{abstract}

We present a microscopic model for a strongly repulsive electron gas
on a 2D square lattice. We suggest that nearest neighbor Coulomb
repulsion stabilizes a state in which electrons undergo a "somersault"
in their internal spin-space (spin-flux). When this spin-1/2
antiferromagnetic (AFM) insulator is doped, the charge carriers
nucleate mobile, charged, bosonic vortex solitons accompanied by
unoccupied states deep inside the Mott-Hubbard charge-transfer
gap. This model provides a unified microscopic basis for (i)
non-Fermi-liquid transport properties, (ii) mid-infrared optical
absorption, (iii) destruction of AFM long range order with doping,
(iv) angled resolved spectroscopy (ARPES), and (v) d-wave preformed
charged carrier pairs. We use the Configuration Interaction (CI)
method to study the quantum translational and rotational properties of
such pairs. The CI method systematically describes fluctuation and
quantum tunneling corrections to the Hartree-Fock approximation and
recaptures essential features of the (Bethe ansatz) exact solution of
the Hubbard model in 1D.  For a single hole in the 2D AFM plane, we
find a precursor to spin-charge separation. The CI ground state
consists of a bound vortex-antivortex pair, one vortex carrying the
charge and the other one carrying the spin of the doping hole.

\end{abstract}

\narrowtext \twocolumn

\section{Introduction}

\begin{figure}
\centering
\parbox{0.45\textwidth}{\psfig{figure=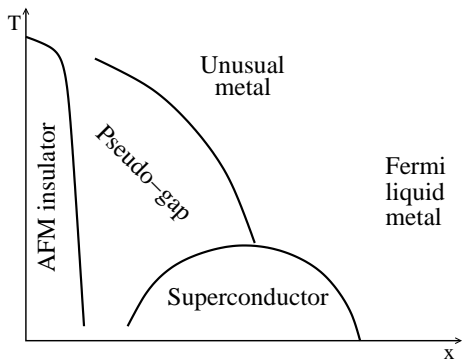,width=85mm,angle=270}}
\caption{\label{fig1} Schematic phase diagram as a function of doping
of high-temperature cuprate compounds. }
\end{figure}

In 1986 Bednorz and Muller \cite{Bed} discovered that the perovskite
(BaLa)$_2$CuO$_4$ exhibits high-temperature superconductivity, with a
critical temperature of up to 30K. Soon after, La$_{2-x}$Sr$_x$CuO$_4$
and YBa$_2$Cu$_3$O$_{7-x}$ were found to have superconducting critical
temperatures of 35K and 95K respectively. Since then, many such
compounds were found, including the Tl and Hg series. The current
record $\mathrm{T_c}$ of 135K (165K under pressure) is found in the
HgBa$_2$Ca$_2$Cu$_3$O$_8$ system. A typical phase diagram is shown in
Fig. \ref{fig1}. The undoped parent is an insulator with long range
antiferromagnetic order. Extremely low doping ( $ x \approx 0.02$
charge carriers per site) leads to a complete destruction of the
long-range AFM order, and a transition to an unusual non-Fermi-liquid
metal. This unusual metal becomes superconducting, with the transition
temperature $\mathrm{T_c}$ strongly dependent on the doping $x$. The
maximum $\mathrm{T_c}$ is reached for dopings around $x=0.15$. For
higher dopings the critical temperature decreases to zero, and in the
overdoped region a crossover towards a (non-superconducting)
Fermi-liquid takes place.

The effective two-dimensional Hamiltonian we use to describe the
electrons residing in the O(2p$_{\sigma}$)-Cu(3d$_{x^2-y^2}$) orbitals
of the isolated CuO$_2$ plane is the generalized one-band Hubbard
Hamiltonian:
\begin{equation}
\label{I1}
{\cal H}= -t \sum_{i,j,\sigma}^{} \left( t_{ij} c_{i\sigma}^{\dagger}
c_{j\sigma} + h.c. \right) + \sum_{i,j}^{} V_{ij} n_i n_j
\end{equation}
where $c_{i\sigma}^{\dagger}$ creates an electron (in the orbital
centered) at site $i$ with spin $\sigma$, $t_{ij}$ is the hopping
amplitude from site $j$ to site $i$ on the square lattice, $n_i\equiv
\sum_{\sigma}c_{i\sigma}^{\dagger}c_{i\sigma}$ is
the total number of electrons at site $i$, and $V_{ij}$ is the Coulomb
repulsion between electrons at sites $i$ and $j$.  The dominant
terms are the nearest-neighbor hopping $t_{ij}=t_0$ and the on-site
Coulomb repulsion $V_{ii}=U/2$. If only these two terms are
considered, and we shift the chemical potential by $U$, this reduces
to the widely studied Hubbard model.

In the undoped parent compound phase, there is one electron per
orbital, and in the absence of interactions 
one would expect these compounds to be metallic, with a
half-filled conduction band. Instead, from Fig. \ref{fig1} we see that
they are insulators with long-range AFM order. This is a
strongly-correlated electron system, with a large on-site repulsion
term ( $U > t$), which causes electrons to become localized one per
each orbital in order to avoid energetically expensive double
occupancy. This obviously leads to insulating behavior, while the
antiferromagnetism is simply a perturbational effect from the (small)
hopping term.\cite{And-J} However, as the plane is doped with holes,
some of the previously filled orbitals are emptied, and electrons in
nearby orbitals can freely hop into them, allowing for charge
conduction. This is also seen from the phase diagram in
Fig.\ref{fig1}, which shows that a very small amount of doping $x
\approx 0.02$ completely destroys the LR AFM insulator phase, and the
system becomes metallic (or superconductor, at lower
temperatures). However, this metal has very unusual non-Fermi liquid
properties. Further doping leads to a crossover to a more
conventional (but non-superconducting) Fermi liquid metal in the
extremely overdoped region.

Understanding the non-Fermi liquid metal above the superconducting
state is the central issue in the cuprate physics. Some of the most
striking evidence of non-Fermi-liquid behavior is provided by Angle
Resolved Photo-Emission Spectroscopy (ARPES), which clearly shows the
absence of quasi-particle peaks in the normal state. Equally
compelling evidence is provided by resistivity measurements,
\cite{resist} which reveal a scattering rate inversely proportional to
the temperature $\tau \sim 1/T$, extending over a range of up to
700K.\cite{49} In an ordinary Fermi-liquid, electron-electron
scattering gives a $T^2$ dependence of the relaxation rate, related to
the quadratic energy dependence of the quasiparticle lifetime ${1
\over \tau }\sim |\epsilon-\epsilon_F|^2$. This is a hallmark of a
Fermi liquid.\cite{AschMer} In fact, the canonical $T^2$ behavior is
indeed observed in the extremely overdoped region, already identified
as a Fermi-liquid. \cite{timusk} But its absence in the intermediate
doping region of the unusual metal, combined with the absence of
quasiparticle peaks in the ARPES data clearly show that this unusual
metal is not a Fermi-liquid.  Then, a natural question arises. If the
charge carriers of the unusual metal are not the quasiparticle-like
charge carriers of a Fermi-liquid, what is their nature?

Some clues are provided by experiments. Hall measurements in the
underdoped regime tell us that the charge carriers are positively
charged and their density equals the doping, i.e. the hole
density.\cite{Hall} (Deviations of both the sign and the density
scaling are observed for some compounds in the overdoped
regions). Optical measurements reveal the appearance, with doping, of
a low-frequency Drude tail, which suggests the existence of free, or
very mobile charge carriers. \cite{timusk} These optical measurements
confirm both the anomalous $1/T$ scattering rate of these charge
carriers, as well as the fact that their density equals the
doping. Together, these measurements suggest that each hole introduced
in the CuO$_2$ layer evolves into a mobile, positively charged,
carrier (but not a quasiparticle). However, both types of measurements
 show other anomalous behavior as well. The Hall coefficient has a
strong, $1/T$ temperature dependence, which is very puzzling, since
neither the density nor the charge of the carriers is temperature
dependent, and neither is the Lorentz force. The optical measurements
reveal that the Drude tail is accompanied by the appearance of a
broad, mid-infrared band deep inside the Mott-Hubbard
gap.\cite{timusk} This mid-infrared band develops with doping, and
signifies the existence of localized levels inside the Mott-Hubbard
gap. Unlike the Drude tail which collapses below $\mathrm{T_c}$ to a
$\delta(\omega)$ function, this mid-infrared response persists
unchanged both above and below $\mathrm{T_c}$. Magnetic measurements
indicate that although the LR AFM is completely suppressed above $x
\approx 0.02$, strong short-range AFM correlations persist, with a
correlation length roughly equal to the average distance between the
holes (charge carriers).\cite{Birg1} Neutron scattering indicates the
appearance of incommensurate peaks in the magnetic structure factor,
with a shift from the AFM $(\pi/a, \pi/a)$ vector varying linearly
with doping for $0.02 < x < 0.12$ and then saturating. The four
incommensurate peaks are arranged diagonally for $ x < 0.05$ and then
rotate by 45$^o$ for $x > 0.05$. \cite{Birg2} The doping dependence of
all these features clearly indicates that magnetism is crucial to the
entire phenomenology.

More puzzling behavior appears in the superconducting state. Flux
quantization clearly proves that pairing does take place, and the unit
of charge in the superconducting condensate is $2e$. ARPES and phase
sensitive measurements have shown that the superconducting gap has
$d$-wave symmetry, and this is believed to mirror the symmetry of the
``Cooper-pair'' wave-function to internal rotations. Penetration depth
measurements show that the density of superfluid in the limit $T
\rightarrow 0$ scales with doping. This means that the
``Cooper-pairs'' must, in fact, be formed from the positive charge
carriers of the unusual metal, not from electrons, as in conventional
BCS theory (the density of electrons is $1-x$, not $x$).  This
underscores the need to identify the charge carrier of the unusual
metal, but leads to another question, namely, how does strong pairing
(leading to high superconducting temperatures) occur in a system
dominated by strong Coulomb repulsion?

In this paper we describe a microscopic model which offers simple and
compelling answers to the above questions, as well as to other
puzzling features described above. Unlike other approaches which
assume that the fundamental quantum degrees of freedom of the
many-electron system are conventional and that the resulting
phenomenology is an ``emergent law of nature'' arising from the
complexity of the system, we propose that there is a {\em hidden
fundamental law of Nature}. This fundamental law of nature expresses a
novel dynamical degree of freedom, namely that an electron can perform
a ``somersault'' in its internal space of Euler angles (when
considered as a rigid body of non-zero volume) as the electron
traverses a closed loop in external coordinate space. The result is a
new quantum number for the many-electron system (the eigenvalues of
the spin-flux), which we propose is as fundamental as the existence of
the spin-${1 \over 2}$ itself. We argue that the electronic somersault
(spin-flux) is dynamically generated through electromagnetic
interactions and in particular, the off-diagonal part of the Coulomb
repulsion between electrons. In Section \ref{sec2} we introduce the
spin-flux Hamiltonian, which we propose as the appropriate Hamiltonian
describing the isolated CuO$_2$ plane. In Section \ref{sec3} we apply
Hartree-Fock Approximation (HFA) and the configuration interaction
(CI) method to the spin-flux Hamiltonian and we identify both the
mobile bosonic charge carrier as well as the nature and symmetry of
the strong pairing interaction between such charge carriers. We
provide a physical justification for the accuracy of the CI
approximation and we explicitly demonstrate this by comparison with
the exact solution of the 1D Hubbard model.\cite{mb4}  Finally, in
Section \ref{sec5} we discuss the comparison of our model and its
results with the experimental findings and draw some final
conclusions.

\section{The spin-flux model}
\label{sec2}

 The neglect of the dynamical consequences of longer range Coulomb
interaction ($V_{ij}=0$, if $i\neq j$), in the generalized Hubbard
model of Eq. (\ref{I1}), is based on the assumption of uniform charge
distribution and on the Fermi-liquid theory notion of screening of the
effective electron-electron interaction.  However, Fermi liquid theory
fails to explain many of the crucial features of the high-T$_c$
cuprates. In our description, we include the nearest-neighbor Coulomb
repulsion, which we assume is on the energy scale of $t$. This has
important dynamical consequences in our model and cannot simply be
absorbed into the Madelung constant.  In particular, it leads to the
generation of spin-flux, an entirely new type of broken symmetry in
the many-electron system, which we show leads naturally to bosonic
charge carriers in the form of meron-vortices, non-Fermi-liquid
behavior and a strong attractive pairing force between holes in the
AFM background.

The concept of spin-flux is closely related to the existence of the
spin$-{1 \over 2}$ particles in nature.  The spin of a physical
electron may be regarded as arising from the quantization of a
classical, symmetric spinning top\cite{Rosen} whose kinematical
properties are described by a set of three independent Euler
angles. These Euler angles constitute a continuous manifold, the group
manifold of SO(3). Unlike the manifold S$_2$ (the surface of a unit
sphere) describing the orientation of a classical magnetic moment, the
group manifold of SO(3) is {\em not} topologically simply
connected. According to the axioms of quantum mechanics any physical
wave-function must be everywhere continuous and differentiable. For
this to be satisfied on a simply connected manifold, the wave-function
must be single valued. Consequently the O(3) nonlinear sigma model
describes integer spins. The doubly connected group manifold of SO(3),
however, can accommodate two-valued wave-functions which are everywhere
continuous and differentiable. This leads to half-integer spins. In
order to accommodate spin$-{1 \over2 }$, the O(3) nonlinear sigma model
must be supplemented with a magnetic monopole which is placed at the
center of the sphere S$_{2}$.\cite{1} The charge of this monopole
corresponds to a quantum of the third Euler angle in the
parameterization of the SO(3) group manifold.

\begin{figure}[h]
\centering
\parbox{0.45\textwidth}{\psfig{figure=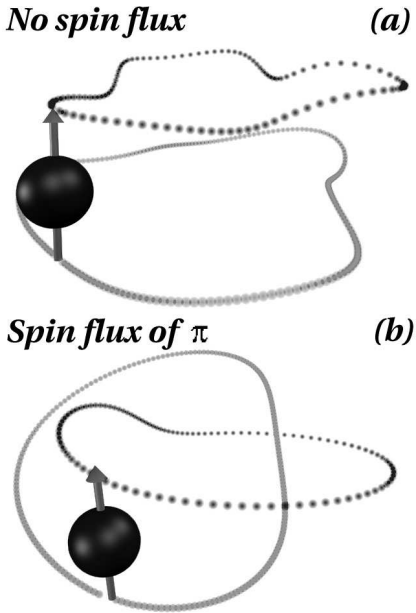,width=55mm}}
\caption{\label{fig2} Conventional (a) vs. spin-flux (b)
trajectory. In the latter case, the spin rotates by $2\pi$
(somersault) as it encircles the path in real space. }
\end{figure}

Spin-flux arises when the electron executes a topologically nontrivial
path within its internal space of Euler angles while it traverses a
closed loop in the external coordinate space.  The internal path is
one that cannot continuously be deformed to zero and corresponds to a
$2\pi$ rotation (somersault) in the space of Euler angles. Another
depiction of this process (see Fig.\ref{fig2}) is seen by
considering distinct points within the spinning top and following
their trajectories as the electron executes a closed path in external
coordinate space. In a non-spin-flux circuit, the two trajectories are
unlinked, whereas in a spin-flux circuit (involving a somersault) the
two trajectories are linked. We propose that this novel possibility
represents a hidden but fundamental Law of Nature which has not been
considered in conventional treatments of many-body theory. Spin-flux
corresponds to a fundamentally new quantum number in a many-electron
system and requires the extension of the conventional many-electron
Hilbert space. We suggest that this simple addition (at a fundamental
level) to the dynamical degrees of freedom of interacting electrons an
a two-dimensional lattice leads to a unified, microscopic explanation
of a large variety of experiments on the cuprates.

In order to describe the above physics from our starting Hamiltonian,
\begin{equation}
\label{1.103}
{\cal H}\!=\!-t_0\!\! \sum_{\!\!\langle i,j\rangle \sigma}^{} \left(
c^{\dagger}_{i\sigma} c_{j\sigma} \!+\! h.c.\!\right)\! +\! U
\!\sum_{i}^{} \!n_{i\uparrow} n_{i\downarrow}\!+\! V \!\!\sum_{\langle
i,j\rangle}^{} n_{i} n_{j}
\end{equation}
we introduce bilinear combination of electron operators $
\Lambda^\mu_{ij}\equiv
c^{\dagger}_{i\alpha}\sigma^\mu_{\alpha\beta}c_{j\beta}$, $\mu=0, 1,
2, 3$, for $i\neq j$ (summation over multiple indexes is assumed).
Here $\sigma^0$ is the $2\times 2$ identity matrix and $\vec
\sigma\equiv (\sigma^1, \sigma^2,\sigma^3)$ are the usual Pauli spin
matrices. The notation $\langle i, j \rangle$ means that the sites $i$
and $j$ are nearest neighbors.  The quantum expectation value
$\langle\ \rangle$ of the $\Lambda^{\mu}_{ij}$ operators are
associated with charge-currents ($\mu=0$) and spin-currents
($\mu=1,2,3$).  Non-vanishing charge currents lead to appearance of
electromagnetic fields, which break the time-reversal symmetry of the
Hamiltonian. Experimentally, this does not occur in the cuprates.  In
the following, we adopt the ansatz that there is no charge current in
the ground state $\Lambda^{0}_{ij}= 0$ but circulating spin-currents
may arise and take the form $\Lambda^{a}_{ij} = {2t_0\over V}
i\Delta_{ij}\hat n_a, a= 1,2,3$, where $\vert \Delta_{ij}\vert
=\Delta$ for all $i$ and $j$, and $\hat{n}$ is a unit vector. These
spin-currents provide a transition state to the uniform spin-flux
mean-field that we use in this paper. In principle, non-uniform states
of spin-flux may arise, in which $\vert \Delta_{ij}\vert$ has a
nontrivial dependence on $i$ and $j$. One such case was discussed
earlier, \cite{1} in which skyrmion textures in the AFM background
carry quantized units of spin-flux. In this case $\Delta_{ij}$ is a
dynamical variable. However, for the purpose of this paper, we
consider only a uniform, static, mean-field configuration of the
spin-flux.

Using the Pauli spin-matrix identity, ${1\over2}
\sigma^\mu_{\alpha\beta}(\sigma^\mu_{\alpha^\prime\beta^\prime})^ \ast
=\delta_{\alpha\alpha^\prime} \delta_{\beta\beta^\prime}$, it is
possible to rewrite the nearest-neighbor electron-electron interaction
terms as $n_in_j=2n_i -{1 \over 2}
\Lambda^\mu_{ij}(\Lambda_{ij}^\mu)^+$. We neglect fluctuations in the
spin-currents, and use the mean-field factorization to replace
$\Lambda^\mu_{ij}(\Lambda^\mu_{ij})^+\rightarrow \langle
\Lambda^\mu_{ij}\rangle (\Lambda^\mu_{ij})^+ +\Lambda^\mu_{ij} \langle
\Lambda^\mu_{ij}\rangle^* -\langle\Lambda^\mu_{ij}\rangle \langle
\Lambda^\mu_{ij}\rangle^*$.  Thus, the quartic nearest-neighbor
Coulomb interaction term is reduced to a quadratic term that is added
to the hopping term leading to the effective Hamiltonian:
\begin{equation}
\label{1.104}
{\cal H}=-t\sum_{\langle i,j \rangle \atop \alpha\beta}^{} \left (
c^{\dagger}_{i\alpha} T^{ij}_{\alpha\beta} c_{j\beta} + h.c. \right) +
U \sum_{i}^{} n_{i\uparrow} n_{i\downarrow}.
\end{equation}
 Here, $T^{ij}_{\alpha\beta}\equiv (\delta_{\alpha\beta}
+i\Delta_{ij}\hat{n}\cdot\vec \sigma_{\alpha\beta})/\sqrt{1+\Delta^2}$
are spin-dependent $SU(2)$ hopping matrix elements defined by the
mean-field theory, and $t=t_o\sqrt{1+\Delta^2}$.  In deriving
(\ref{1.104}) we have dropped constant terms which simply change the
zero of energy as well as terms proportional to $\sum\limits_in_i$
which simply change the chemical potential. It was shown previously
\cite{1,2} that the ground state energy of the Hamiltonian of
Eq.(\ref{1.104}) depends on the SU(2) matrices $T^{ij}$ only through
the plaquette matrix product $T^{12}T^{23}T^{34}T^{41}\equiv\exp
(i\hat n\cdot\vec\sigma\Phi)$.  Here, $\Phi$ is the spin-flux which
passes through each plaquette and $2\Phi$ is the angle through which
the internal coordinate system of the electron rotates as it encircles
the plaquette.  Since the electron spinor wave-function is two-valued,
there are only two possible choices for $\Phi$.  If $\Phi=0$ we can
set $T^{ij}_{\alpha\beta}=\delta_{ij}$ and the Hamiltonian
(\ref{1.104}) describes conventional ordered magnetic states of the
Hubbard model.  The other possibility is that a spin-flux $\Phi=\pi$
penetrates each plaquette, leading to
$T^{12}T^{23}T^{34}T^{41}=-1$. This means that the one-electron
wave-functions are antisymmetric around each of the plaquettes,
i.e. that as an electron encircles a plaquette, its wave-function in
the internal spin space of Euler angles rotates by $2\pi$ in response
to strong interactions with the other electrons.  In effect, the
electron performs an internal ``somersault'' as it traverses a closed
path in the CuO$_2$ plane. \cite{1} This spin-flux phase is
accompanied by a AFM local moment background (with reduced magnitude
relative to the AFM phase of the conventional Hubbard model). In the
spin-flux phase, the kinetic energy term in (\ref{1.104}) exhibits
broken symmetry as though a spin-orbit interaction has been added.  In
the presence of charge carriers this mean-field is unstable to the
proliferation of topological fluctuations (magnetic solitons) which
eventually destroy AFM long range order. In this sense, the analysis
which we present below goes beyond simple mean field theory. The
quantum dynamics of these magnetic solitons described by the
Configuration Interaction (CI) method, corresponds to tunneling
effects not contained in the Hartree-Fock approximation.  For
simplicity, throughout this paper we assume that the mean-field
spin-flux parameters $T^{ij}$ are given by the simplest possible
choice $T^{12}=-1, T^{23}=T^{34}=T^{41}=1$ (for more details, see
Ref. 14). In order to go beyond a mean-field description of
the spin-flux, these matrices may also be treated as dynamical
variables. In this paper, we go beyond mean-field theory in describing
the antiferromagnetic degrees of freedom but restrict ourselves to a
mean-field model of the spin-flux.

\begin{figure}
\centering
\parbox{0.45\textwidth}{\psfig{figure=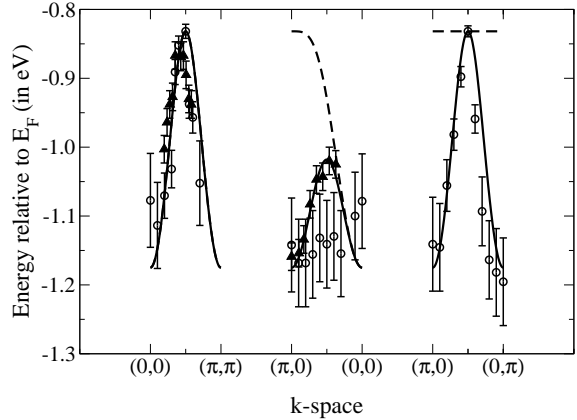,width=85mm,angle=270}}
\caption{\label{fig3} A comparison between the experimentally
determined $E(\vec{k})$ quasi-particle dispersion relation, from angle
resolved photoemission studies (ARPES), for the insulating
Sr$_2$CuO$_2$Cl$_2$ and the HF AFM spin-flux model dispersion relation
(full line) and the HF AFM conventional Hubbard model dispersion
relation (dashed line).  Three directions in $\vec{k}-$space
are shown: $(0,0)$ to $(\pi,\pi)$, $(\pi,0)$ to $(0,0)$ and $(\pi,0)$
to $(0,\pi)$.  While the peak on the $(0,0)$ to $(\pi,\pi)$ is equally
well described in both models, the mean-field spin-flux model gives a
much better agreement for the $(\pi,0)$ to $(0,0)$ and $(\pi,0)$ to
$(0,\pi)$ directions.  The fitting corresponds to $U=2.01$ eV,
$t=0.29$ eV for the spin-flux phase, and $U=1.98$ eV, $t=0.21$ eV in
the conventional phase. The experimental results are the ARPES results
of Ref. 15 (circles) and Ref. 16 (triangles). }
\end{figure}

The mean-field ground-state of the undoped spin-flux model is an AFM
Mott insulator.  It is interesting to note that the quasi-particle
dispersion relation obtained in the presence of the
spin-flux\cite{mb3} accurately recaptures the dispersion as measured
through angle-resolved photo-emission spectroscopy (ARPES) in a
compound such as Sr$_2$CuO$_2$Cl$_2$ \cite{Wells} (see
Fig. \ref{fig3}).  There is a a large peak centered at $(\pi/2,
\pi/2)$ with an isotropic dispersion relation around it, observed on
both the $(0,0)$ to $(\pi,\pi)$ and $(0,\pi)$ to $(\pi,0)$ lines. The
spin-flux model in HFA exhibits another smaller peak at $(0,\pi/2)$
which has been observed in more recent experimental data.\cite{newAR}
The quasi-particle dispersion relation of the conventional Hubbard
model ($T^{12}=T^{23}=T^{34}=T^{41}=1$) has a large peak at
$(\pi/2,\pi/2)$ on the $(0,0)$ to $(\pi,\pi)$ line (see
Fig. \ref{fig3}), but it is perfectly flat on the $(0,\pi)$ to
$(\pi,0)$ line (which is part of the large nested Fermi surface of the
conventional 2D Hubbard model).  Also, it has a large crossing from
the upper to the lower band-edge on the $(0,0)$ to $(0,\pi)$
line. Both this dispersion relation and the very similar one of the
$t-J$ model (see Ref. 15) are in contradiction to ARPES
data.

\section{Doping Induced Meron-Vortex Solitons}
\label{sec3}

\subsection{ The Static Hartree-Fock Approximation}

The HF results for the undoped AFM ground state of the spin-flux
Hamiltonian are in good agreement with experimentally measured
dispersion (see Fig. \ref{fig3}).  The azimuthal symmetry of the
dispersion relations about the Fermi points plays a key role in
determining the symmetries of doping induced magnetic
configurations. This is more straightforward to see in a simpler,
continuum version of the model, obtained by letting the lattice
constant $a \rightarrow 0$ (see Ref. 17,18). Since the
dispersion relation near the Fermi point $\vec q = ({\pi\over
2a},{\pi\over 2a})$ is isotropic, the dependence on $\vec k = \vec K -
\vec q \rightarrow -i \nabla_{\vec r}$ of the continuum HF equations
is such that it preserves rotational invariance. As a result, the 2D
HF equation reduces trivially to a 1D radial equation, with a
structure very similar to that of the 1D differential HF equation
obtained for the 1D Hubbard model.\cite{mb1,mb2} Once this radial 1D
solution is found, the 2D configuration is simply generated through a
$2\pi$ rotation about an axis perpendicular to the 2D plane. As a
result, there is a close analogy between solutions obtained for the 1D
Hubbard model and for the 2D spin-flux model, in all our
investigations.\cite{mb4,mb1,mb2,mb5}

\begin{figure}
\centering
\parbox{0.45\textwidth}{\psfig{figure=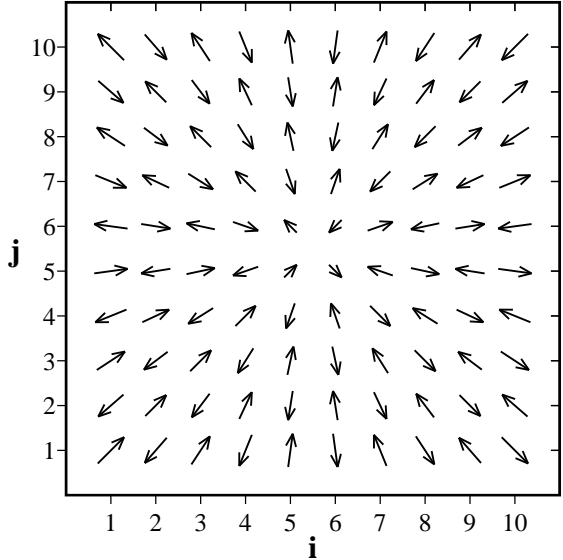,width=85mm}}
\caption{\label{fig4} Self-consistent spin  distributions of
a 10x10 lattice with a meron-vortex in the spin-flux phase. The core
of the meron is localized in the center of a plaquette, in the
spin-flux phase (in the conventional phase, the core of the
meron-vortex is localized at a site). This excitation has a
topological winding number 1, since the spins on either sublattice
rotate by $2\pi$ on any curve surrounding the core. The magnitude of
the staggered magnetic moments is slightly diminished near the vortex
core but is equal to that of the undoped AFM background far from the
core. The hole is localized on in the vortex core.}
\end{figure}

In the corresponding discrete model, the relevant doped configuration
is the meron-vortex (see Fig.\ref{fig4}).  The doping hole is trapped
in the core of a magnetic vortex, which indeed has azimuthal
symmetry. The bound level on which the hole is trapped can be
shown\cite{mb2,mb3} to be split from the top of the valence band and
drawn deep inside the Mott-Hubbard gap. As a result, the meron-vortex
is a charged boson. This can be inferred by direct inspection of
Fig.\ref{fig4}, which shows a configuration with total spin zero and a
positive charge trapped in its core. An argument based on the
electronic structure, identical to the one offered for the charged
bosonic domain-walls of the polyacetylene, also
holds.\cite{mb1,mb2,mb3,pol-rev} The parallel to the quasi-one
dimensional 1D polyacetylene is again a reflection of azimuthal
symmetry which reduces the 2D continuum model to a 1D radial equation.
Clearly, the isotropic dispersion about the Fermi points is
responsible for the appearance of {\em bosonic} charge carriers. They
are very unlike quasi-particle charge carriers, which carry both spin
and charge together. Thus, one would expect a metal with such bosonic
charge carriers to have intrinsically different properties from those
of a Fermi-liquid metal.

\begin{figure}
\centering
\parbox{0.45\textwidth}{\psfig{figure=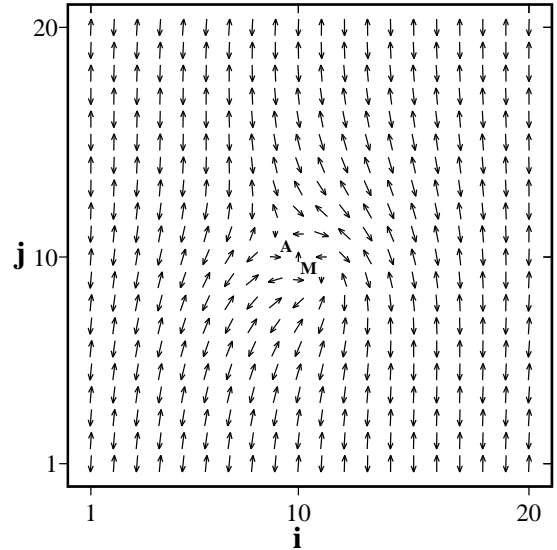,width=85mm}}
\caption{\label{fig5} Self-consistent spin distribution for
a tightly bound meron-antimeron pair. The meron (M) and the antimeron
(A) are localized on neighboring sites. The total winding number of
the pair is zero. The magnetic AFM order is disturbed only on the
region where the vortices are localized.  The attraction between holes
is of topological nature and on long length scale is stronger than
unscreened Coulomb repulsion between charges.  The doping charge is
mostly localized on the two plaquettes containing the meron and
antimeron cores. The two holes localized in the vortex cores are
responsible for the fact that the meron-antimeron pair does not
collapse.  }
\end{figure}

A look at the spin-configuration in Fig.\ref{fig4} also shows that
this cannot be realized if cyclic boundary conditions (CBC) are
imposed. With them, the self-consistent solution found in the presence
of one hole is a very different configuration called the spin-bag,
which is a rather immobile quasiparticle-like configuration (carrying
both charge $+e$ and spin-${1 \over 2}$).\cite{mb3,mb5} However, for
more than one hole added to the AFM plane, the HF ground-state of the
spin-flux Hamiltonian always shows meron-vortices created through
doping, even with CBC.\cite{mb3} The simple reason for this is that
while a single isolated meron-vortex is incompatible with CBC,
meron-antimeron pairs are not (see Fig. \ref{fig5} for a typical
meron-antimeron configuration). In fact, the nucleation of merons and
antimerons in pairs also solves another problem, related to the
energetic cost of creating a meron. It is straightforward to prove
that the energy of a single meron configuration increases
logarithmically with its size.\cite{mb3} Beyond a certain (large)
separation it is energetically favorable for the meron-antimeron pair
to collapse into a pair of charged spin bags. In the conventional
Hubbard model (with no spin-flux) spin bags are favored at all
separations. In the spin-flux phase, the farther the meron is from the
antimeron, the more spins in between are rotated by the vortices, and
the excitation energy increases. As a result, an isolated pair tends
to be as closely bound as possible. If the vortices were uncharged, at
low temperatures they would annihilate each other. However, the holes
localized in the vortex cores lead to a very strong short-range
Coulomb repulsion which prevents the pair annihilation, thus making
the pair stable. It is worth noting that even in the complete absence
of screening, at long distances the $1/r$ Coulomb repulsion would be
overcome by the $\ln(r)$ attraction between vortices, leading to a
stable bound pair.

To conclude, we see that even at the static HF level, charge carriers
in the spin-flux phase exhibit bosonic nature, and a strong pairing
attraction to other charge carriers. This attraction is of magnetic
origin, arising from exchange energy lost by spins which are no longer
perfectly AFM aligned. The importance of the bosonic meron-vortex
excitations becomes even more apparent when we consider fluctuation
and tunneling corrections to the HF approximation. These correspond
to translational motion of the charged vortices. It turns out that
charged meron-vortices have an effective mass comparable to that of
the band electron. As seen from the CI method, they are much more
mobile than the very heavy spin-bags.

\subsection{Fluctuations and Tunneling:  The Configuration
Interaction Method}

We now discuss how to improve the mean-field description above. Given
a complete basis of states $\{|\phi_i\rangle\}$ spanning the $N$-body
Hilbert space, the exact $N$-body ground state wave-function can be
written as $|\Psi \rangle = \sum_{i}^{} \alpha_i |\phi_i\rangle$. The
coefficients $\{\alpha_i\}$ are found from solving the Schr\"{o}dinger
equation ${\cal H} |\Psi\rangle = E | \Psi \rangle$, which reduces to
a linear system of equations $ \sum_{j}^{}{\cal H}_{ij} \alpha_j = E
\sum_{j}^{} {\cal O}_{ij} \alpha_j$, for all $i$. Here, ${\cal
H}_{ij}=\langle \phi_i| {\cal H} | \phi_j\rangle $ and ${\cal
O}_{ij}=\langle \phi_i| \phi_j\rangle $ are the matrix elements of the
Hamiltonian in the complete basis chosen and the overlapping matrix
for the basis states, respectively.

In general, the number of basis states of the $N$-body problem
increases exponentially with $N$, and the problem becomes untractable
for values of $N$ which are rather small. To deal with large values of
$N$, one is forced to truncate the complete basis set, and only retain
a smaller subset of states which are most likely to contribute
substantially in the decomposition of the ground-state wave-function.
The choice of this smaller subset is the crucial issue, since it
determines the quality of the approximation. In what follows, we
describe a particular subset which allows us to recapture certain key
features of the exact Bethe ansatz solution of the 1D Hubbard model.

Consider the set of Slater determinants generated from the
 Hartree-Fock solution. Its states are $|\Psi_{HF}\rangle$ and all
 possible particle-hole excitations $\{ a_h^{\dagger} a_p
 |\Psi_{HF}\rangle\}$, $\{ a_h^{\dagger}a_{h'}^{\dagger} a_p a_{p'}
 |\Psi_{HF}\rangle\}$ etc.  Obviously, if all possible combinations of
 occupied and empty orbitals are considered, the set thus spanned is a
 complete basis of the $N$-body space.

Let us now order the states in this HF basis set according to their
energies $\langle \phi | {\cal H}| \phi \rangle$. If we are interested
in the ground-state and the low-lying excitations of the system, we
only need to keep the low-energy states of the HF basis set.  This
procedure is, in fact, very well known for Hamiltonians with a {\em
non-degenerate} HF ground-state.  If only the HF ground-state and the
states with one particle-hole excitations are kept, this leads to the
Random Phase Approximation (RPA). Besides a better approximation for
the ground-state than simple HFA, the RPA enables us to find
collective excitations and the particle-hole continuum.

\begin{figure}[t]
\centering
\parbox{0.45\textwidth}{\psfig{figure=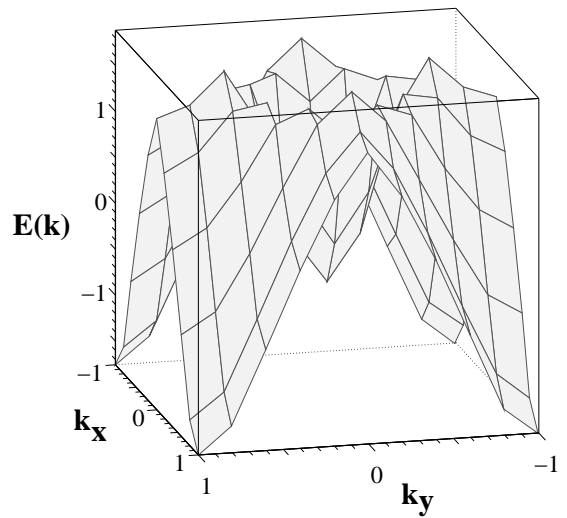,width=75mm}}
\caption{\label{fig6} The lowest energy dispersion band $E_J(\vec{k})$
(in units of $t$) as a function of the total momentum $\vec {k}$ of
the meron-antimeron pair. The momentum units are $\pi/a$ and $U/t=5$.
For convenience, the reference energy is taken to be the static HF
energy of the self-consistent meron-antimeron pair.  Quantum hopping
and rotation lowers the overall energy of the pair by $1.75t$.  }
\end{figure}

On the other hand, the doped HF ground-states of the spin-flux
Hamiltonian are {\em degenerate}. For instance, the meron-antimeron
pair shown in Fig.\ref{fig5} happens to be centered at site
(10,10). It is obvious that configurations which have the
meron-antimeron pair centered about any other site 
will have the exact same HF energy (cyclic boundary conditions are
imposed). Also, if the center for the pair is fixed at a site, there
are four distinct possible arrangements of the meron and antimeron
about that site, obtained by rotating the meron-antimeron axis by
90$^o$. Thus, for a plaquette of size $N_x \times N_y$, the HF ground
state is $4N_xN_y$ degenerate.  Clearly, the minimal choice for the
subset of states used to search for the ground-state of the doped
system must include all these degenerate HF ground-state
wave-functions. This choice is the essence of the configuration
interaction (CI) approximation.\cite{spanish} One expects these HF
states to mix with equal weight $|\alpha_i|^2$ in the decomposition of
the ground-state and low-energy states $|\Psi\rangle =
\sum_{i=1}^{4N_xN_y} \alpha_i |\phi_i\rangle$.  Specifically, we
denote $|\phi_i\rangle \rightarrow |\Psi_{\theta}(n,m)\rangle$ to be
the HF ground-state wave-function describing a configuration centered
at site $(n,m)$ of the lattice, and with the meron-antimeron axis at
an angle $\theta$ from the $x$-axis. Then, $1 \le n \le N_x$, $1\le m
\le N_y$ and $\theta=45^o, 135^o, 225^o $ or $315^o$. On general
symmetry grounds one expects the ground-state and the low-lying energy
states to have the general form
\begin{equation}
\label{3.1}
|\Psi_J(\vec k) \rangle = \sum_{n, m, \theta}^{} e^{i (k_xn +k_ym)a}
 e^{i J \theta} |\Psi_{\theta}(n,m)\rangle
\end{equation}
The cyclic boundary conditions limit the vector $\vec k$ to a subset
of equally spaced values inside the first Brillouin zone and $J$ must
be an integer.

\begin{figure}
\centering
\parbox{0.45\textwidth}{\psfig{figure=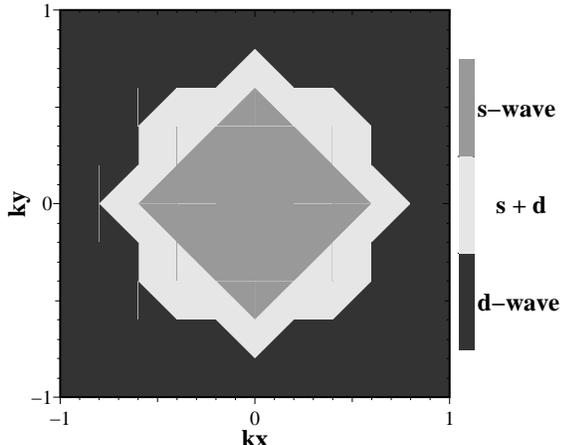,width=75mm}}
\caption{\label{fig7} The rotational symmetry of the meron-antimeron
wave-function as a function of the total momentum carried by the pair
(measured in units of $\pi/a$).  The outside region (containing the
absolute minima points $(\pi,\pi)$) has d-wave symmetry ($J=2$), while
the core region about the $(0,0)$ point has s-wave symmetry
($J=0$). The intermediary area is a mix of s+d wave symmetry.}
\end{figure}

Eq.(\ref{3.1}) shows that the ground-state and low-lying energy
states found within the CI method have translational and rotational
symmetry. This procedure overcomes the most glaring shortcoming of the
mean-field theory (the broken translational and rotational
invariance). Clearly, the CI wave-functions describe quantum dynamics
of the charge carriers.  The pair is no longer pinned at one site, as
in HFA, but moves over the entire lattice. The mobility of the pair
and its preferred internal angular momentum can be obtained from the
dispersion relation $E_J(\vec k) = \langle \Psi_J(\vec k) | {\cal H} |
\Psi_J(\vec k)\rangle $. The lowest energy band of the meron-antimeron
pair obtained for $U/t=5$ is shown in Fig.\ref{fig6}. \cite{mb5} We
find that the CI ground-state corresponds to pairs of total momentum
$(\pi/a,\pi/a)$, and which have $d$-wave symmetry, $J=2$. In fact, $J$
varies throughout the Brillouin zone as shown in Fig.\ref{fig7}, from
pure $d$-wave around the $(\pm\pi/a,\pm\pi/a)$ points to pure $s$-wave
around the $(0,0)$ point, where a second local minimum exists. The
large width ($\approx 4t$) of the dispersion band clearly proves that
the meron-antimeron pair is a very mobile excitation. The existence of
mobile charged bosonic merons and antimerons thus provides a
microscopic basis for the non-Fermi liquid ``parent'' metal from which
superconductivity emerges.

\begin{figure}
\centering
\parbox{0.45\textwidth}{\psfig{figure=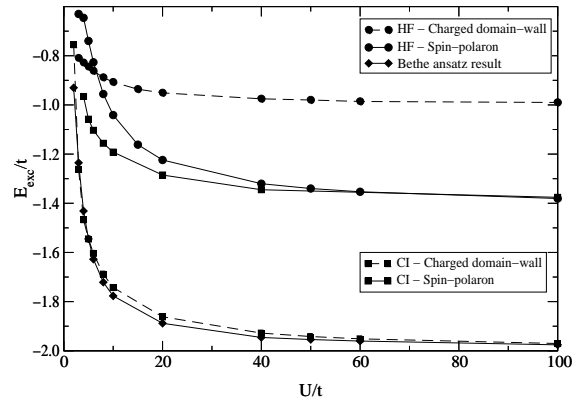,width=85mm,angle=270}}
\caption{\label{fig8} Excitation energy, in units of $t$, for a
mobile charged domain wall (squares, dashed line) 
and a mobile charged spin polaron
(squares, full line), as obtained from the CI approach. The exact excitation
energy given by the Bethe-ansatz method is shown by diamonds. The
domain-wall CI energy is in excellent agreement with the exact BA
results overt the entire $U/t$ range, while the spin polaron CI energy is
significantly different. For comparison, we also show the excitation
energies for the domain wall (circles, dashed line) and the
spin-polaron (circles, full line)  as obtained within 
HFA, proving again that the extra kinetic
energy gained by the moving domain wall is of order $t$ for most $U/t$
values. In contrast, the extra kinetic energy gained by the
spin polaron is of order $t^2/U \rightarrow 0$ as $U/t$ increases, so
in the large $U/t$ limit there is almost no difference between the HF
and CI results for the charged spin polaron. We conclude that the
charged domain-wall is the relevant excitation for all values of
$U/t$.}
\end{figure}

As an indication of the validity and accuracy of the CI method, we
briefly review the results of such an analysis of the 1D Hubbard
model.\cite{mb4} Here, the analog of the a charged meron-vortex is the charged
domain wall soliton which facilitates a $\pi$ flip from one AFM
ground-state to the other AFM ground-state (with all spins
flipped). 
 The charged domain-wall traps the hole on a pair of levels
that are localized deep inside the Mott-Hubbard gap, and is a
charged boson.\cite{mb1,mb2,mb4} The analog of the spin-bag is the
spin-polaron, which disturbs the AFM order only locally, trapping the
hole in its small FM core. As a result, the spin-polaron carries both
charge and spin$-{1 \over2 }$. In Fig.\ref{fig8} we show the HF
(circles) and  CI (squares) energies of both the spin-polaron (full
line) and the charged domain-wall (dashed line). For the
spin-polaron, the difference between the HF and CI energy varies as
$t^2/U$. In analogy with the 2D case, this is the expected behavior
since the spin$-{1 \over 2}$  carried by the spin-polaron localizes
it on one magnetic sublattice. On the other hand, the domain-wall
lowers its energy by an amount of the order $t$ between the HF and the
CI values. This clearly shows that the charged domain-wall is a very
mobile objects, and that  its dynamics must be properly described in
order to get a realistic picture. When this is done, it is apparent
that the charged domain-wall is the low energy excitation over the
entire $U/t$ range, and in fact its energy is in excellent agreement
with the exact Bethe Ansatz prediction (shown as diamonds).

Finally, we consider the stability of the ground-state and the
low-lying energy states to inclusion of more states in the subset used
to generate them.  The next meaningful enlargement of this subset is
to add to it all states with one particle-hole excitation, obtained
from all degenerate HF ground-state wave-functions. While this
increases the size of the subset substantially, it does not lead to a
lower ground-state energy. This is because the system is gapped.  The
lowest energy particle-hole excitation is obtained when an electron
from the top of the valence band is excited on one of the bound
levels, localized in the cores of the meron and the antimeron. These
levels are roughly one quarter of the way into the gap (for $U/t=5$),
so the states with a particle hole excitation have an energy of order
$US$ more than the HF ground-states. Consequently they do not mix into
the ground-state. They only influence the higher energy states. The
low-lying excitations depicted in Fig. \ref{fig6} are {\em charge}
excitations, associated with motion of the charged meron-antimeron
pair. These are not affected by addition of excited states in the
variational subset, except at rather high energy.  But there also
exist low-lying {\em spin-wave} excitations, which are described by
adding particle-hole excited states (RPA) in the variational subset.
While it is very likely that the nucleation of merons and antimerons
in the AFM background will alter the dispersion of the spin-waves, it
is interesting to note that low-lying charge and low-lying spin
excitations have very different origin.  This is related to the
spin-charge separation tendency exhibited by this system.

If a single hole is added to the 2D AFM plane and the cyclic boundary
conditions are imposed, the HF approximation leads to a spin-bag
mean-field solution for the spin-flux phase. We analyze the
translational properties of the spin-bag using the CI method. As it
turns out,\cite{mb5} is that the spin-bag is a very immobile object.
The width of its dispersion band is of the order of $t^2/U$, i.e. much
smaller than the $\sim t$ bandwidth of the meron dispersion. The
reason is that the spin-bag carries both charge and spin-${1 \over
2}$. The spin-${1 \over 2}$ only allows the spin-bag to live on one of
the magnetic sublattices. (If it were to move to the other sublattice,
its spin should flip, and such processes are forbidden). In order to
move, a spin-bag must tunnel two sites to a second-nearest neighbor
site, and this is a $t^2/U$ process. On the other hand, the charged
meron carries no spin, so there are no restrictions for its motion. It
can move to a nearest neighbor site, leading to a $t$-hopping
process. This suggests that when dynamics is properly taken into
consideration within CI, a highly mobile vortex-antivortex pair
sharing a single hole is energetically favored to the immobile
spin-bag. We have verified this hypothesis, and demonstrated that an
upper bound to the energy of a singly-charged vortex-antivortex pair
is, indeed, much lower than the CI energy of spin-bag.\cite{mb5} The
charged spin-bag solution is thus unstable to dissociation into a
highly mobile vortex-antivortex pair, which shares the hole. When a
second hole is added, a second single-charged vortex-antivortex pair
is nucleated. However, it becomes energetically more convenient for
the two holes to become bound to the same vortex-antivortex pair,
leading to the appearance of a meron-antimeron pair which carries both
charges (pre-formed D-wave pair).  The remaining uncharged
vortex-antivortex pair is unstable to collapse, at low temperatures.

\section{Discussion and comparison with experiments}
\label{sec5}

The crucial distinguishing feature of our model is the concept of
spin-flux, the dynamical possibility of an electron undergoing a
somersault as it traverses a closed loop.  In the undoped parent
compound, this leads to dispersion relations with isotropic symmetry
about the Fermi point, in excellent agreement with those measured
experimentally through ARPES. This symmetry of the dispersion
relations leads to real-space configurations which have the same type
of symmetry. In the spin-flux model, the holes doped into the AFM
plane nucleate magnetic vortices and become trapped in their cores,
leading to the appearance of mobile, bosonic, charge carriers.

In contrast, both the conventional Hubbard model and its asymptotic
limit, the t-J model, exhibit a very large, nested Fermi surface (at
the mean-field level) in the undoped parent compound. This Fermi
surface has quasi-linear 1D character, since there is no
$k$-dependence along the Fermi surface. As a
result, configurations stabilized by doping exhibit the same quasi-1D
character in real space. It has been suggested they take the form of
charged stripes. \cite{stripes} We note finally that vortex-like
configurations are unstable in the conventional Hubbard model, while
stripe-like configurations are generally unstable to the formation of
a quantum liquid of merons in the spin-flux model. One notable
exception is the commensurate case $x=0.125$, when merons and
antimerons crystallize along 1D lines, leading to the stripe-like
configuration observed experimentally,\cite{Tranquada} provided that a
small (3\%) hopping anisotropy is included in the model. \cite{mb3}

The mobile bosonic charged meron-vortices created through doping
provide a microscopic basis for a non-Fermi liquid behavior. They also
exhibit a very strong pairing attraction, of magnetic origin. This
pairing, which arises in a purely repulsive electron system, leads to
appearance of pre-formed ``Cooper-pairs'' of $d$-wave symmetry.  This
agrees with the experimental findings of $d$-wave superconductivity
and the scaling of the superfluid density with doping. The pre-formed
pairs may also be related to the observation of a pseudo-gap on the
underdoped side of the phase diagram. \cite{pseudo,preformed}

Many other features of our model are in agreement with experimentally
observed properties of the cuprate superconductors. Nucleation of
magnetic vortices with doping explains a variety of magnetic
properties, starting with complete destruction of the long-range AFM
order for very low doping concentration. As we can see from
Fig. \ref{fig5}, a tightly-bound meron-antimeron pairs disturbs the
long-range AFM ordering of about 100 spins of the lattice. For very
low dopings, these pairs are far from each other, and there are many
spins on the plane whose orientations are not affected by any
pair. Thus, most of the spins maintain the long range AFM
order. However, as the doping increases over about $2 \%$, the areas
occupied by each meron-antimeron pairs start to overlap with those
occupied by the neighboring pairs. At this doping the orientation of
all the spins on the CuO$_2$ planes is affected by at least one pair
of vortices, and therefore the LRO is lost. The local ordering,
however, is still AFM. This picture explains the extremely low doping
necessary for the disappearance of LR AFM order, as well as the fact
that the spin correlation length is basically equal to the average
distance between holes (vortices) and does not depend strongly on the
temperature.\cite{Birg1} Each hole carries its vortex with it, and the
spins in each vortex are correlated with each other. The correlation
length is thus roughly equal to the average inter-vortex (inter-hole)
distance.  The nucleation of magnetic vortices quantitatively explains
the split of the $(\pi,\pi)$ AFM Bragg peak into the four
incommensurate peaks whose positions shift with doping,\cite{Birg2} as
observed in LaCuO and, more recently, in YBaCuO.\cite{arai} The form
factor of any given vortex already gives rise to an apparent splitting
of the neutron scattering peak.  As demonstrated in Ref. 14,
even at the mean-field level we recapture the neutron scattering data
using the HF distribution of meron-vortices. A more detailed
investigation suggests that saturation\cite{Aeppli} of the peak
splitting for $x > 0.12$ may be related to expansion of the core
radius of the individual vortices at higher doping.\cite{mb6}

Optical behavior of the cuprates is also explained naturally using our
model. Two features develop in the optical absorption spectra with
doping: a broad mid-infrared temperature-independent absorption band,
and a strongly temperature-dependent low-frequency Drude
tail. \cite{timusk} In our model the broad mid-infrared band is
related to excitation of electrons from the valence band onto the
empty levels bound in the vortex cores, \cite{mb3} which are localized
approximately one quarter of the way inside the Mott-Hubbard gap. The
number of localized levels scales with the number of vortices, and
inter-vortex and spin-wave interactions lead to their broadening into
the observed band. This mechanism is similar to the one leading to a
broad mid-infrared absorption band in polyacetylene with
doping. \cite{Tanaka} (In the continuum limit, the meron vortex in
fact creates a pair of mid-gap states in the Mott-Hubbard
gap\cite{mb1}).  The polyacetylene band is due to electronic
excitations inside the cores of the polyacetylene domain-wall
solitons,\cite{pol-rev} which are the topological analogues of
meron-vortices.\cite{mb1,mb2} Another strong argument in favor of this
interpretation is provided by photoinduced absorption
experiments. \cite{foster} If the undoped parent compounds are
illuminated with intense visible light, they develop absorption bands
that resemble the mid-infrared bands of the doped compounds. Similar
behavior is observed in polyacetylene, and is attributed to the
nucleation of solitons by photoexcited electron-hole
pairs. \cite{Orenstein}

 The second component of the optical spectrum is the Drude tail. It
results from the response of the freely moving charged vortices to the
external electric field. The strong temperature dependence of this
tail is determined by the scattering mechanism for merons (due to
interactions with other merons and spin-waves). This interpretation is
also supported by the fact that the superconducting transition leaves
the mid-infrared absorption band unchanged.  Merons with internal
electronic structure are still present on the planes but pair
condensation leads to a collapse of the Drude tail into a
$\delta(\omega)$ response.

Finally, our model provides some understanding of the cross-over
towards the Fermi-liquid metal in the overdoped phase.  For large
dopings ($\delta > 0.30-0.40$) the average inter-vortex spacing
becomes extremely small and the very cores of the merons start to
overlap. In this limit the Mott-Hubbard gap is completely filled in by
the discrete levels, and the spin-flux state becomes energetically
unstable relative to a normal Fermi liquid. \cite{mb3}

We note, finally, that the spin-flux Hamiltonian has essentially no
free or adjustable parameters. The choice of $U/t$ is fixed by the
experimentally measured size of the Mott-Hubbard charge transfer gap
at zero doping. All other experimental features such as (i) the
position and nature of the mid-infrared optical absorption band, (ii)
the ARPES data and (iii) the position of the magnetic neutron
scattering satellite peaks as a function of doping, are quantitatively
described by the same choice.  More detailed comparisons with specific
experiments may require the incorporation of specific (smaller energy
scale) interactions which are not included in this simplest version of
the spin-flux Hamiltonian.

\section*{Acknowledgments}

M.B. acknowledges support from a Natural Sciences and Engineering
Research Council of Canada Postdoctoral Fellowship.

\end{document}